\documentclass[journal]{IEEEtran}
\ifCLASSINFOpdf
\else
  \usepackage[dvips]{graphicx}
  \usepackage[export]{adjustbox}
\fi
\usepackage[cmex10]{amsmath}
\usepackage{mathabx}
\begin{document}
%
\title{Out-of-Band Radiation Comparison of GFDM, WCP-COQAM and OFDM at Equal Spectral Efficiency}
%
%
%

\author{Ali~Bulut~\"{U}\c{c}\"{u}nc\"{u},~\IEEEmembership{Student Member,~IEEE},
        and Ali~\"{O}zg\"{u}r~Y{\i}lmaz,~\IEEEmembership{Member,~IEEE}
\thanks{The authors are with the Department of Electrical and Electronics Engineering, Middle East Technical University, Ankara, Turkey (e-mail: ucuncu@metu.edu.tr, aoyilmaz@metu.edu.tr)}}

%
%

\markboth{IEEE Signal Processing Letters, Submitted Draft 2015}%
{Shell \MakeLowercase{\textit{et al.}}: Bare Demo of IEEEtran.cls for Journals}
%



\maketitle

\begin{abstract}
GFDM and WCP-COQAM are amongst the candidate physical layer modulation formats to be used in 5G, whose claimed lower out-of-band (OOB) emissions are important with respect to cognitive radio based dynamic spectrum access solutions. In this study, we compare OFDM, GFDM and WCP-COQAM in terms of OOB emissions in a fair manner such that their spectral efficiencies are the same and OOB emission reduction techniques are applied to all of the modulation types. Analytical PSD expressions are also correlated with the simulation based OOB emission results. Maintaining the same spectral efficiency, carrier frequency offset immunities will also be compared.
\end{abstract}


%
\IEEEpeerreviewmaketitle

\section{Introduction}
\IEEEPARstart{T}{he} emergence of 4G telecommunication standards and its applications in commercial standards have enabled data rates of several hundreds of megabits/s. As the physical layer modulation technique in those standards, OFDM is used owing to its robustness against multipath distortion \cite{nee2000ofdm}. However, filter bank multi-carrier (FBMC) schemes compete with orthogonal frequency division multiplexing (OFDM), due to several reasons, for 5G and other future systems. One particular reason is that pulse shaping can be applied to have high spectral containment in FBMC, whereas high spectral OOB leakage occurs for OFDM because of its rectangular pulse shape \cite{farhang2011ofdm}. Moreover, higher spectral efficiency by omitting the orthogonality requirement in pulse design and lower complexity in uplink scenarios are some of the other advantages of FBMC compared to OFDM \cite{farhang2011ofdm}, \cite{barbieri2009time}.


Amongst the FBMC methods, a popular one is the generalized frequency division multiplexing (GFDM) proposed in \cite{GFDM_5G}. A variant of GFDM called WCP-COQAM, which employs OQAM type modulation, is proposed in \cite{lin2014multi}. A significant advantage of GFDM and WCP-COQAM is claimed to be their lower OOB radiation compared to OFDM \cite{GFDM_5G},~\cite{lin2014multi}. This is in accordance with the aforementioned potential of lower OOB radiation in FBMC schemes.

 
In all of the aforementioned OOB emission comparisons, GFDM or WCP-COQAM is compared to OFDM under different spectral efficiency conditions. However, we believe that a fair comparison with respect to OOB emissions should be made when the spectral efficiencies of the three modulation types are made equal. The details of equating their spectral efficiencies will be provided in the following sections of the paper. On the other hand, these schemes should also be compared in terms of carrier frequency offset (CFO) immunity for the equal spectral efficiency case, the reason of which will be clear in the subsequent parts of the paper. Furthermore, in the existing comparisons between OFDM and GFDM or WCP-COQAM in literature, windowing and guard symbols insertion is applied only to GFDM or WCP-COQAM. In this study, windowing and guard symbol insertion will also be applied to OFDM in the OOB emission comparisons in order to be fair.

The organization for the remainder of the article is as follows. First, frame structures of the three modulation types will be presented. Next, the signal models for the three modulation schemes will be stated. After that, some analytical power spectral density (PSD) expressions will be provided to relate them with the OOB emission results. Finally, the results concerning OOB emissions and CFO immunity performances will be presented for the three modulation techniques under unequal or equal spectral efficiency conditions.

\section{GFDM Frame structure}
\begin{figure}[htbp]
\hspace{24pt}\includegraphics[width=0.65\columnwidth,left]{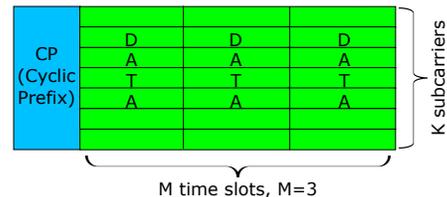}\vspace{-10pt}
\caption{GFDM frame structure ($M=3$ time slots, $K$ subcarriers)}
\label{fig:GFDM_frame}
\end{figure}

\begin{figure}[htbp]
\hspace{22pt}
\includegraphics[width=0.85\columnwidth,left]{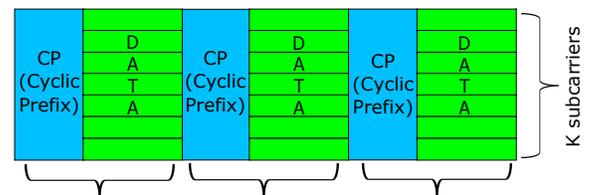}\vspace{-10pt}
\caption{3 consecutive OFDM symbols with $K$ subcarriers}
\label{fig:OFDM_frame}
\end{figure}

GFDM is a special multicarrier scheme with a structure that includes cyclic prefix (CP) allowing block equalization at the receiver, similar to OFDM. However, GFDM is different from OFDM in that there can be more than one time slots in a GFDM symbol, whereas there is only one in OFDM. To illustrate this, an OFDM frame consisting of three OFDM symbols and a GFDM frame consisting of $M=3$ time slots with $K$ subcarriers is presented in Fig.~\ref{fig:GFDM_frame} and Fig.~\ref{fig:OFDM_frame}. \vspace{-10pt}
 \vspace{-20pt}

As can be inferred from Fig.~\ref{fig:GFDM_frame} and Fig.~\ref{fig:OFDM_frame}, three consecutive OFDM symbols with $K$ subcarriers should be transmitted in order to transmit the same number of data symbols as a single GFDM symbol (or a GFDM frame, two terms are used interchangeably in this paper) with $K$ subcarriers and $M=3$ time-slots. Since for each OFDM symbol, there is a CP overhead, 3 CPs should be transmitted in OFDM, whereas only a single CP is enough for GFDM. From this point, GFDM seems to be spectrally more efficient than OFDM. However, it is also possible to increase the number of subcarriers of OFDM to $M \cdot K$ keeping the same bandwidth. This will result in an longer OFDM symbol in time that has the same duration as a GFDM symbol which corresponds to packing $M\cdot K$ subcarriers into the same bandwidth. In this case the spectral efficiencies of OFDM and GFDM will be the same. Similar arguments also hold between OFDM and WCP-COQAM frames since the frame structures of GFDM and WCP-COQAM are the same.

\section{Signal Model}
\label{sec:Signal_model}
The transmitted signal for a single GFDM symbol with $M$ time slots and $K$ subcarriers can be written as \cite{GFDM_5G}
\begin{equation}
x[n]=\sum_{k=0}^{K-1}\sum_{m=0}^{M-1}d_{k,m} p[(n-mK)_{MK}] e^{-j2\pi \frac{kn}{K}}
\label{eq:GFDM_SignalModel}
\end{equation}
where $n = 0,\ldots,MK-1$, $d_{k,m}$ corresponds to the data symbol transmitted at the $k^{th}$ subcarrier and $m^{th}$ time slot in the GFDM symbol. The number of time slots and subcarriers in a GFDM symbol are $M$ and $K$, respectively. Moreover, $p[(n-mK)_{MK}]$ denotes circular shift of the pulse shaping function $p[n]$, which is of length $MK$, by $mK$ with modulo $MK$. This equation can also represent OFDM, when $M=1$ and $p[n]$ has a rectangular shape. For WCP-COQAM, the transmitted signal can be expressed as \cite{lin2014multi}
\begin{equation}
\label{eqn:WCP-COQAM transmitted waveform}
\begin{split}
\hfill x&_{WCP-COQAM}[n]= \\
&\sum_{k=0}^{K-1}\sum_{m={0}}^{2M-1} \tilde{d}_{k,m}p[(n-mK/2)_N]e^{j2\pi\frac{k}{K}(n-D/2)}e^{j\phi_{k,m}}
\end{split}
\end{equation}
where $\tilde{d}_{k,m}$ are real valued data symbols, as opposed to complex valued data symbols in GFDM or OFDM and $D=MK-1$ \cite{lin2014multi}. For the signal model in (\ref{eqn:WCP-COQAM transmitted waveform}) the windowing function is neglected for simplicity. Moreover, there is also the $e^{j\phi_{k,m}}$ term for which there is no unique selection \cite{Siohan2002_May}. One possible selection is $e^{j\phi_{k,m}}=\frac{\pi}{2}(k+m)$ \cite{lin2014multi}, \cite{Siohan2002_May}. As can be inferred from (\ref{eqn:WCP-COQAM transmitted waveform}), the real valued data symbols are transmitted with an offset of $K/2$ samples, which characterizes the effect of OQAM type modulation.

\section{Analytical Expressions for PSD}
As mentioned before, the spectral efficiencies of OFDM, GFDM and WCP-COQAM will be equated by inserting more subcarriers into the same bandwidth in OFDM. This is possible if the duration of the OFDM symbol should be increased. This results in the sinc functions in the frequency spectrum of an OFDM signal to decay faster, which will have a decreasing effect on the OOB emissions. On the other hand, since there will be more subcarriers, the OOB portion of the PSD of OFDM signal will be composed of the summation of a larger number of sinc functions, which will result in higher OOB emissions. Therefore, it is not very straightforward to conclude whether the OOB emissions will increase or decrease as a result of the aforementioned two effects. However, we can directly use analytical expressions to estimate the effect of increasing the number of subcarriers while keeping the same bandwidth on the OOB emissions. For rectangular pulse shaping, power spectral density (PSD) of the OFDM signal can be written as \cite{OFDM_analytical_PSD}
\begin{equation}
\label{eqn:PSD_sinc_superpos}
P_x(f)=\frac{\sigma_c^2}{LT_s}\sum_{k=0}^{K-1}(sinc_L[(f-k\Delta_f)LT_s])^2,
\end{equation}
where $P_x(f)$ is the PSD of the OFDM signal, $\sigma_c^2$ is the variance of the data symbols, $T_s=1/F_s$ is the sampling period, $K$ is the number of subcarriers, $\Delta_f=\frac{1}{KT_s}$ is the frequency spacing between the subcarriers and $L$ is the total number of samples in OFDM symbol including the time guard interval (for cyclic prefix or zero-padding). The $sinc_L(f)$ is the aliased sinc function defined as
\begin{equation}
\label{eqn:aliased_sinc_expr}
sinc_L(f)=\begin{cases}
(-1)^f(M-1), & \text{if} f\in Z \\
\dfrac{sin(\pi Mf)}{Msin(\pi f)}, & \text{if} f\notin Z \\
\end{cases}
\end{equation}
If we increase the number of subcarriers, $K$, to $MK$, where $M$ is the number of time slots in GFDM frame, the new PSD expression for OFDM becomes
\begin{equation}
\label{eqn:PSD_sinc_superpos_eq_spec_eff}
P_{x,eq}(f)=\frac{\sigma_c^2}{L'T_s}\sum_{k=0}^{MK-1}(sinc_{L'}[(f-k\Delta_f/M)L'T_s])^2,
\end{equation}
where $L'=M(L-N_{guard})+N_{guard}$ is the total number of the samples in OFDM symbol with $M\cdot K$ subcarriers, where $N_{guard}$ is the number of samples in the time guard interval. Note that $L'>L$. The mentioned effects of increasing the number of subcarriers on PSD can also be observed by comparing (\ref{eqn:PSD_sinc_superpos}) and (\ref{eqn:PSD_sinc_superpos_eq_spec_eff}). More sinc terms are summed in (\ref{eqn:PSD_sinc_superpos_eq_spec_eff}) compared to (\ref{eqn:PSD_sinc_superpos}), which has an increasing effect on the PSD values at OOB frequencies, whereas the $L'T_s$ multiplier in the sinc term in (\ref{eqn:PSD_sinc_superpos_eq_spec_eff}) being larger than $LT_s$ in that of (\ref{eqn:PSD_sinc_superpos}) causes a faster decay for each sinc term. Furthermore, the 1/L' multiplier in (\ref{eqn:PSD_sinc_superpos_eq_spec_eff}) will also decrease the PSD values, but its effect vanishes if the maximum values of the PSDs are normalized to 0 dB. In fact, such a normalization ensures the same average transmitted power. The overall change in the PSD values can be calculated using (\ref{eqn:PSD_sinc_superpos}) and (\ref{eqn:PSD_sinc_superpos_eq_spec_eff}) to foresee the effect of increasing the number of subcarriers in OFDM while keeping the same bandwidth.
\section{Simulation Results}
\subsection{Unequal Spectral Efficiency Conditions}
\label{sec:Unequal Spectral Efficiency Conditions}
To see the OOB radiation for OFDM, GFDM and WCP-COQAM the parameters in Table~\ref{table:sim_param} are used. Number of subcarriers, guard subcarriers and cyclic-prefix (CP) length is chosen from the possible choices specified in the LTE standard~\cite{LTE_standard}. When guard subcarriers are used in OFDM and GFDM they will be denoted as G-OFDM, G-GFDM, respectively. Windowing is also applied to reduce OOB emissions. Windowed versions will be represented as W-OFDM and W-GFDM (windowing is present by default in WCP-COQAM) and the results for which both windowing and guards symbols are used will be referred to as GW-OFDM, GW-GFDM and GWCP-COQAM. RC pulse is used for GFDM and WCP-COQAM. However, since it is a non-orthogonal pulse, discrete Zak transform (DZT) based orthogonalization methods in \cite{Bolcskeidzak} is applied to RC pulse to be used with WCP-COQAM to yield both time-frequency well localized and orthogonal pulse causing zero inter-symbol interference (ISI) and inter-carrier interference (ICI).
\begin{table}[t]
\renewcommand{\arraystretch}{1.2}%
\caption{Simulation Parameters}
\label{table:sim_param}
\begin{center}
\begin{tabular}{  l | c  }
\hline
Parameter  & Value\\
\hline
Total number of subcarriers (K) & 128 (Part A) or 1152 (Part B)    \\
No. of guard subcarriers  & 52 or 684 \\
Pulse Shape     	      & RC (with roll-off=0.1)    \\
Constellation             & 4-QAM\\
Number of symbols (M)     & 9     \\
CP length				  & 32     \\
Windowing				  & Hanning, 18 samples from both sides \\
Spectral estimation method& Periodogram \\
Interpolation filter type & RC pulse with roll of 0.1 \\
Interpolation filter duration & 81 symbols \\
No. of Monte-Carlo simulations& 300 \\
 
\hline
\end{tabular}
\end{center}
\end{table}

\begin{figure}[htbp]
\centering
\includegraphics[width=\columnwidth]{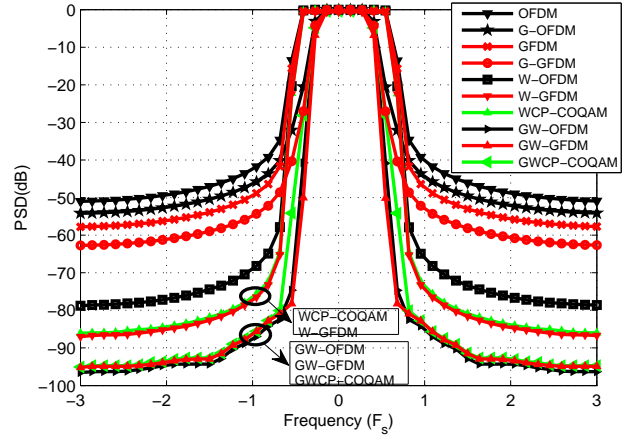}
\caption{PSD for OFDM, GFDM and WCP-COQAM under different spectral efficiency conditions}
\label{fig:OOB_128OFDM}
\end{figure}

One other important point is how oversampling is done to see the frequency ranges larger than the transmission bandwidth. In our simulations, time-domain signals are sixfold oversampled using an RC filter of length 81 symbols. However, after filtering, the samples are truncated from both sides, in order that the total number of samples is 6 times the number of samples in the signals that are not oversampled. Otherwise, depending on the length of the interpolation filter, OFDM, GFDM or WCP-COQAM symbols would leak to the neighbouring symbols, which will require additional cyclic prefix length that will decrease the overall spectral efficiency. Furthermore, if such a truncation is not made, interpolation filter itself would give the effect of windowing, which can convey results that windowing has little or no effect in terms of OOB emissions. Using the aforementioned parameters and methods the OOB emissions for unequal spectral efficiency case are obtained as in Fig.~\ref{fig:OOB_128OFDM}.  Note that the maximum value of the PSDs are normalized to 0 dB. Moreover, the frequency axis is normalized with respect the the sampling frequency $F_s$.

As can be seen on Fig.~\ref{fig:OOB_128OFDM}, there is about 9 dB advantage of GFDM over OFDM or G-GFDM over G-OFDM in terms of the PSD values at $f=\pm 3F_s$. When windowing is applied, W-GFDM and WCP-COQAM has about 7 dB lower PSD values compared to W-OFDM. These results promote the use of GFDM or WCP-COQAM in terms of OOB emissions as also pointed out in \cite{GFDM_5G}, \cite{lin2014multi}. However, when both guard symbol insertion and windowing is applied, which is the common practice, the PSD values for the three modulation types are the same, which can be seen from the PSD curves named as GW-OFDM, GW-GFDM and GWCP-COQAM in Fig.\ref{fig:OOB_128OFDM}. Therefore, it can be stated that, for unequal spectral efficiency conditions specified, the OOB emissions of OFDM, GFDM and WCP-COQAM are similar when both windowing and guard symbol insertion techniques are used. However, when neither windowing nor guard symbol insertion is used, there is an advantage for GFDM or WCP-COQAM compared to OFDM.

\subsection{Equal Spectral Efficiency Conditions}
\begin{figure}[htbp]
\centering
\includegraphics[width=\columnwidth]{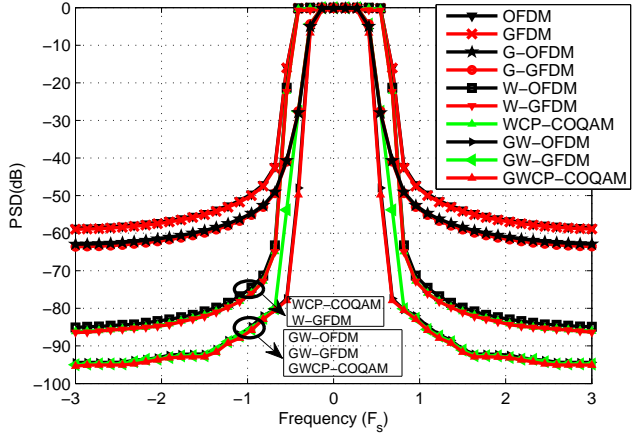}
\caption{PSD for OFDM, GFDM and WCP-COQAM under equal spectral efficiency conditions}
\label{fig:OOB_1024OFDM}
\end{figure}

To equate the spectral efficiencies, the number of subcarriers in OFDM is increased to $M \cdot K$ as stated before. This means that $ 128 \times 9=1152$ subcarriers should be transmitted with OFDM. In LTE standard, although a transmission with such a number of subcarriers are not supported, the ratio of the number of occupied and total subcarriers in unequal spectral efficiency case, which is 76/128, is maintained to have 684 (76/128*9=684) occupied subcarriers in OFDM. For GFDM or WCP-COQAM, the transmission parameters are selected to be the same as in Part~\ref{sec:Unequal Spectral Efficiency Conditions}. For these parameters, OOB emissions are observed as in Fig.~\ref{fig:OOB_1024OFDM}.

As can be observed from Fig. \ref{fig:OOB_1024OFDM}, the PSD values for the three modulation types are the same whether guard symbol insertion and/or windowing is applied or not. Furthermore, the PSD values for OFDM at $f=\pm 2F_s$ for unequal or equal spectral efficiency cases are calculated by replacing $L=128+32$, $K=128$, $M=9$, $L'=1152+32$ in (\ref{eqn:PSD_sinc_superpos}) and (\ref{eqn:PSD_sinc_superpos_eq_spec_eff}) which has yielded a PSD difference about 7.8 dB at $f=\pm 2F_s$. In fact, this value is obtained when the multiplier terms $\sigma_c^2/LT_s$ or $\sigma_c^2/(L'T_s)$ in (\ref{eqn:PSD_sinc_superpos}) and (\ref{eqn:PSD_sinc_superpos_eq_spec_eff}) are taken to be equal since maximum PSD values obtained from the simulations are normalized to 0 dB. This value (7.8 dB) can also be observed when the PSD values of OFDM without windowing or guard symbol insertion in Fig. \ref{fig:OOB_128OFDM} and Fig. \ref{fig:OOB_1024OFDM} at $f=\pm 2F_s$ are compared. Therefore, the OOB emission change for OFDM between unequal and equal spectral efficiency cases that are expected from (\ref{eqn:PSD_sinc_superpos}) and (\ref{eqn:PSD_sinc_superpos_eq_spec_eff}) is consistent with the Monte-Carlo based simulation results.

\begin{figure}[htbp]
\centering
\includegraphics[width=\columnwidth]{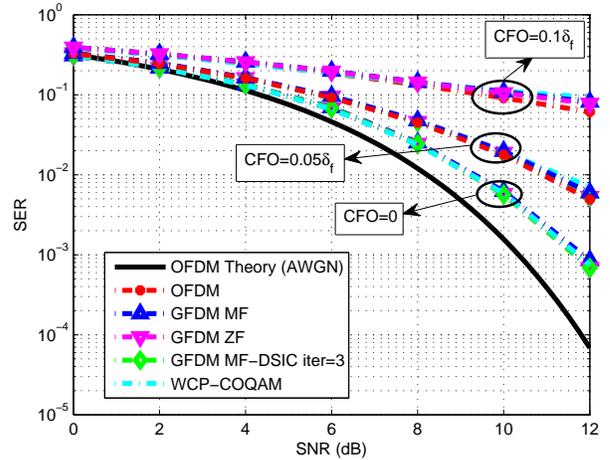}
\caption{SER v.s SNR for OFDM, GFDM and WCP-COQAM with parameters in Table~\ref{table:sim_param} under equal spectral efficiency conditions}
\label{fig:SER_SNR}
\end{figure}

At this point, one may ask whether the CFO immunity for OFDM will be compromised, when the number of subcarriers in OFDM are increased to ensure the same spectral efficiency. This corresponds to packing more subcarriers into the same bandwidth for OFDM, which results in an increased CFO sensitivity. However, for GFDM or WCP-COQAM, although there are a smaller number of subcarriers compared to OFDM, the ISI between the time slot symbols within a GFDM or WCP-COQAM symbol will also be increased under CFO. Therefore, it is not very straightforward to make a conclusion in terms of the performance degradations of the two modulation schemes under CFO. Towards this end, with the parameters used in equal spectral efficiency case in OOB simulations, uncoded symbol error rate (SER) vs signal to noise ratio (SNR) performances in the static ISI COST-207 hilly terrain model channel \cite{proakis2007digital} for OFDM, GFDM and WCP-COQAM are obtained as in Fig.~\ref{fig:SER_SNR}. Perfect channel knowledge is assumed and single-tap zero forcing type equalization is performed at the receiver side. Different GFDM receiver types, namely matched filter receiver (GFDM-MF) \cite{GFDM_5G}, zero-forcing (GFDM-ZF) \cite{GFDM_5G}, matched filter followed by an interference cancellation receiver (GFDM-MF-DSIC) \cite{alvesperformance} with three iterations are used. The analytical AWGN performance of OFDM (indicated as OFDM-AWGN in Fig.\ref{fig:SER_SNR}) is borrowed from \cite{bahai2004multi}. Moreover, $\delta_f$ represents the subcarrier spacing in Fig. \ref{fig:OOB_1024OFDM}.

Observed from Fig.~\ref{fig:SER_SNR}, as CFO increases, OFDM does not perform worse than the other modulation schemes with different receiver types. This shows that although the number of subcarriers in OFDM are significantly higher than GFDM or WCP-COQAM, OFDM does not exhibit a worse CFO immunity performance compared to GFDM or WCP-COQAM, since the ISI between the time slots in GFDM or WCP-COQAM also occur owing to CFO. This validates the fairness of the comparison between OFDM, GFDM and WCP-COQAM in terms of their OOB radiation levels under the same spectral efficiency conditions. One may also note that the conclusions that we draw will change according to the selection of the pulse shape. With the same simulation parameters, we have obtained similar results for RRC (with roll-off 0.1 or 0.3), Dirichlet or Gaussian pulses in terms of the OOB emission and CFO immunity performances of the three modulation schemes.

\section{Conclusion}
In this work, 5G candidate modulation schemes, GFDM and WCP-COQAM are compared to OFDM in terms of OOB radiation levels. Unlike the ones in literature, we make these comparisons under the same spectral efficiency conditions and apply OOB emission suppression techniques to all three modulation types to reach fair comparison grounds. Moreover, we also calculate the effect of changing the number of subcarriers in OFDM to ensure equal spectral efficiency on the PSD of OFDM signal based on analytic PSD expressions. The OOB emission results conveyed that GFDM and WCP-COQAM are superior to OFDM for the unequal spectral efficiency case, although they have the same OOB radiation performance when windowing and guard symbol insertion techniques are applied. However, when their spectral efficiencies are set equal, the OOB emissions of the three modulation schemes are observed to be very similar regardless of the application of windowing and guard symbol insertion techniques. The CFO immunity comparisons for the equal spectral efficiency case further validated the fairness in OOB emission comparisons. In conclusion, despite having higher complexity, GFDM or WCP-COQAM has no significant OOB emission advantage over OFDM under equal spectral efficiency. The same conclusion holds for the unequal spectral efficiency case when standard windowing and guard symbol insertion techniques are utilized.

%
%

\ifCLASSOPTIONcaptionsoff
  \newpage
\fi

\end{document}